\documentclass[preprint,showpacs,preprintnumbers,amsmath,amssymb]{revtex4}

\usepackage{graphicx}

\begin{document}

\title[Non-isolated dynamic black and white holes]{Non-isolated dynamic black
holes and white holes}

\author{M. L. McClure}
\email{mcclure@astro.utoronto.ca}
\author{Kaem Anderson}
\author{Kirk Bardahl}
\affiliation{Mathematics \& Applied Mathematics, University of Cape Town, 
Rondebosch 7701, South Africa}

\date{\today}

\begin{abstract}
Modifying the Kerr-Schild transformation used to generate black and white 
hole spacetimes, new dynamic black and white holes are obtained 
using a time-dependent Kerr-Schild scalar field. 
Physical solutions are found for black holes that shrink with time and for
white holes that expand with time.
The black hole spacetimes are physical only in the vicinity of the black 
hole, with the physical region increasing in radius with time.
The white hole spacetimes are physical throughout. 
Unlike the standard Schwarzschild solution the singularities are 
non-isolated, since the time-dependence introduces a mass-energy 
distribution. 
The surfaces in the metrics where $g_{tt}=g^{rr}=0$ are dynamic, moving 
inward with time for the black holes and outward for the white holes, 
which leads to a question of whether these spacetimes truly have event 
horizons---a problem shared with Vaidya's cosmological black hole 
spacetimes. 
By finding a surface that shrinks or expands at the same rate as the null 
geodesics move, and within which null geodesics move inward or outward 
faster than the surfaces shrink or expand respectively, it is verified 
that these do in fact behave like black and white holes. 
\end{abstract}


\maketitle

\section{Introduction}

Non-isolated dynamic black holes are of interest since they are more 
realistic than black holes that exist by themselves in a vacuum and 
that never evolve in time.
White hole solutions are also of interest: white hole exteriors are 
better representations of stars than black hole exteriors since white 
holes can act as sources of radiation while black holes can only act as 
sinks.
Many dynamic black hole solutions are cosmological black holes, but it 
would be ideal to have solutions for black holes that are dynamic without 
relying on a cosmological background to achieve this.

In this paper new non-isolated dynamic black and white holes will be 
obtained via Kerr-Schild transformations \cite{Ker65} (see also 
\cite{Ste03}). 
A Kerr-Schild transformation 
\begin{equation}
\bar{g}_{ab} = g_{ab} + 2Hl_al_b
\end{equation}
can be used to generate a new metric $\bar{g}_{ab}$ by taking a known 
metric $g_{ab}$ and adding a component based on a scalar field $H$ and 
null geodesic vector field $l^a$.
For example, the Eddington-Finkelstein form of the Schwarzschild metric 
can be derived by performing a Kerr-Schild transformation on Minkowski 
space, with $2H=2m/r$ and $l_a = (1,1,0,0)$ or $l_a = (1,-1,0,0)$ for a 
black hole or white hole respectively.

Previously Dawood and Ghosh \cite{Daw04} found a family of dynamic black 
and white holes by combining the time-dependence of Vaidya's radiating 
star \cite{Vai43} with Salgado's \cite{Sal03} family of static black holes.
Cosmological black holes---such as those of McVittie \cite{McV33},
Vaidya \cite{Vai77}, Thakurta \cite{Tha81}, Sultana and Dyer \cite{Sul05}, 
and McClure and Dyer \cite{McC06}---also provide examples of dynamic 
black holes since the black holes are surrounded by an expanding 
cosmological background and evolve with the universe.
Swiss cheese black holes \cite{Ein45} can also be surrounded by a 
cosmological background; however, they possess a vacuum region between the 
black hole and the surrounding FLRW universe, and the black holes 
themselves are not dynamic.

Cosmological black hole spacetimes are generally obtained either by 
performing the same conformal transformation on the Schwarzschild 
spacetime as is used to transform Minkowski space to FLRW, or by 
performing the same Kerr-Schild transformation on FLRW as is used to 
transform Minkowski space to Schwarzschild.
The essential difference between the two methods is whether the Kerr-Schild 
scalar field contains the cosmological scale factor $R(t)$.
This suggests another possible scenario---one in which the seed metric is 
Minkowski but the Kerr-Schild scalar field contains a scale factor 
$R(t)$, which is the scenario that will be explored in this paper.
This type of spacetime could lead to solutions of dynamic black and white 
holes in an asymptotically-Minkowski background, rather than in a 
cosmological background.
New solutions of Einstein's field equations will be found with this 
spacetime in Section II. 
The Einstein tensors will be calculated using the computer algebra 
program REDUCE 3.8 with the Redten 4.3 package \cite{Har94}.

Ashtekar and Krishnan \cite{Ash04} give a detailed review of the 
various definitions used to describe the horizons of dynamic black holes.
Commonly black holes and white holes are specified by future and past 
event horizons respectively, but for dynamic spacetimes these horizons 
will generally not correspond with the apparent horizons that track the 
surfaces that locally behave like the black and white hole boundaries at 
a moment in time.
Event horizons are defined using boundaries of the 
causal past of future null infinity and causal future of past null 
infinity respectively, but it is more practical to look at trapped 
surfaces.
On a trapped surface the outward-directed null geodesics converge such 
that the volume they occupy is decreasing in time, and on a 
marginally trapped surface their volume expansion is zero.
The boundary of the total trapped region is used to specify an apparent 
horizon or marginally trapped surface that locally acts like the 
boundary of a black hole at a moment in time.
Hayward \cite{Hay94} defines a trapping horizon in terms of marginally
trapped surfaces such that it is essentially the time evolution of an
apparent horizon.
In general, the expansion of the event horizon may be positive, since the
area may be increasing, in which case it is not necessary that the event
horizon coincides with the apparent horizon, or that any trapped surfaces
even exist within the event horizon at a given moment of time 
(see Wald and Iyer \cite{Wal91}).
During black hole collapses, it is expected that as mass is accreted the 
apparent horizon forms and moves outwards, eventually approaching the 
event horizon.
For instance, Schnetter and Krishnan \cite{Sch06} show that the black 
hole version of Vaidya's radiating star (a radiation sink) has an event 
horizon outside the $r=2m$ apparent horizon, such that the apparent 
horizon asymptotically approaches the event horizon as $m$ increases.

It is generally assumed that an apparent horizon cannot exist outside of 
an event horizon.
However, a region that is trapped or marginally trapped could subsequently 
cease to be, since these definitions only specify what a surface is doing 
at a specific moment in time.
In special circumstances, an apparent horizon could move inward and an 
event horizon could only exist if there is an inner region that remains 
trapped for all time. 
For instance, if a conformal transformation is performed to shrink 
the Schwarzschild spacetime, this could lead to photons becoming trapped 
outside the event horizon (which remains at $r=2m$ under conformal 
transformation) since the contraction of the space influences the volume 
expansion of photons within it. 
Assuming the rate at which the space shrinks is decreasing, then regions 
that are trapped due to this negative volume expansion would subsequently 
cease to be, such that the apparent horizon moves inward and ultimately 
coincides with the event horizon.
Thus, with a non-isolated black hole where the gravitational contraction of
matter is creating a negative volume expansion, it is possible for an event 
horizon to exist within an apparent horizon.
If an apparent horizon is shrinking, ultimately it is not a very satisfying 
demonstration of the existence of a black hole: unless there is an inner 
region that remains trapped for all time, then the trapped region 
will shrink to nothing and allow all outgoing null geodesics to 
eventually escape.

In the case of the new dynamic solutions in this paper, $g_{tt}=g^{rr}=0$ at 
$r=2mR^2$, so that the null geodesics remain at fixed areal radius $r$ while 
this surface is itself shrinking or expanding in $r$ with time according to 
the scale factor $R$.
Thus, for shrinking black holes photons may escape the $r=2mR^2$ surface, 
and for expanding white holes photons may become enveloped within the surface, 
meaning the $r=2mR^2$ surface cannot act as an event horizon.
If these are indeed black holes and white holes, there should be a surface 
contained within the $r=2mR^2$ surface where photons are held fixed 
relative to the surface as it shrinks or expands, and within which 
photons move inward or outward faster than surfaces shrink or expand for 
the black holes and white holes respectively. 
These surfaces will be found in Section III.

Vaidya's cosmological black holes \cite{Vai77} can be expressed as
\begin{equation}
\mathrm{d}s^2 = [R(t)]^2 \left(- \mathrm{d}{t}^2 + \mathrm{d}r^2 + r^2
(\mathrm{d}{\theta}^2 + \mathrm{sin}^2 \theta \; \mathrm{d}
{\phi}^2) \right) +
\left(\frac{2m}{r}\right)(\mathrm{d}t+\mathrm{d}r)^2
\end{equation}
for black holes in asymptotically-flat universes.
Vaidya's cosmological black holes also have the problem that the 
$g_{tt}=g^{rr}=0$ surface shrinks while null geodesics are held at fixed $r$ 
at that surface, allowing photons to escape this surface.
Vaidya claimed this surface is an event horizon by first 
performing the calculation for the case of a black hole in a static 
Einstein universe and then extending it to the case of an expanding 
universe, but this surface can no longer serve as an event 
horizon once it becomes dynamic. 
(Also, since Vaidya considered the cases of closed universes, it is not strictly
possible to satisfy the definition of an event horizon in such universes.)
In Section III it will be shown that the new dynamic black hole spacetimes 
are conformal to Vaidya's cosmological black holes in asymptotically-flat 
universes, so if event horizons exist for the new dynamic black holes 
then they must exist for these cosmological black holes as well.

Finally, in Section IV the mass of the new black and white holes will be 
allowed to vary across null surfaces analogous to Vaidya's radiating star 
\cite{Vai43}.
Interpretations will be given for non-isolated dynamic black holes and 
white holes with a perfect fluid component and an additional null-fluid 
component.

\section{New dynamic black hole and white hole solutions}

Performing a Kerr-Schild transformation of Minkowski space with a
scale factor $R(t)$ modifying the usual scalar field used to obtain the 
Schwarzschild metric, the new line element
\begin{equation}
\mathrm{d}s^2 =  - \mathrm{d}t^2 + \mathrm{d}r^2 + r^2 
(\mathrm{d}\theta^2+\mathrm{sin}^2\theta \mathrm{d}\phi^2) +
\frac{2m[R(t)]^2}{r}(\mathrm{d}t \pm \mathrm{d}r)^2 
\end{equation}
is obtained (with the plus or minus signs corresponding to the black 
or white hole cases respectively).
Looking at the Einstein tensor (with $G_{ab}=-\kappa T_{ab}$), the only
non-zero components are
\begin{eqnarray}
G^0_1 = - G^1_0 = \frac{4m\dot{R}R}{r^2} \\
G^1_1 = \mp \frac{8m\dot{R}R}{r^2} \\
G^2_2 = G^3_3 = \frac{2m(\ddot{R}R+\dot{R}^2)}{r}.
\end{eqnarray}
Assuming there is a perfect fluid component in the energy-momentum tensor, 
then the energy density $\mu$ and pressure $p$ are related by
\begin{equation}
G^a_a = \kappa (\mu - 3p) = \mp \frac{8m\dot{R}R}{r^2} + 
\frac{4m(\ddot{R}R+\dot{R}^2)}{r},
\end{equation}
since $G^a_a = 0$ for any additional null fluid or heat conduction 
component that might be present in the energy-momentum tensor. 
From spherical symmetry $G^2_2 = G^3_3 = - \kappa p$, so 
\begin{equation}
p = - \frac{2m(\ddot{R}R+\dot{R}^2)}{\kappa r} 
\end{equation}
and
\begin{equation}
\mu = \mp \frac{8m\dot{R}R}{\kappa r^2} - 
\frac{2m(\ddot{R}R+\dot{R}^2)}{\kappa r}. 
\end{equation}

For the new spacetimes to yield a valid physical solution of Einstein's 
field equations, the energy conditions (e.g.\ see \cite{Wal84}) must be 
satisfied.
With a perfect fluid this essentially requires that $\mu \geq 0$ and $\mu 
+ p \geq 0$ (weak 
energy condition), $\mu + p \geq 0$ and $\mu +3p \geq 0$ (strong energy 
condition), and  $\mu \geq |p|$ (dominant energy condition).

In the black hole case, in order for the energy density to be non-negative 
everywhere, both $\dot{R}R$ and $\ddot{R}R+\dot{R}^2$ must be non-positive; 
however, this is not possible for any $R(t)$.
Looking for a solution in a region of spacetime (with one term of the 
energy density positive and dominating the negative term) requires the 
pressure to be negative; otherwise, the positive term of the energy 
density would be the same as the pressure term, and the negative term in 
the energy density would cause the pressure to be of greater magnitude 
than the energy density, violating the dominant energy condition.
Thus, $\dot{R}R$ must be negative, which requires $R(t) \sim t^x$ with $x 
< 0$, such that the black holes shrink with time.
The positive term of the energy density goes as $t/r$ relative to the 
negative term, so the energy density is positive and all the energy 
conditions are satisfied for sufficiently small values of $r$, and with 
the radius of the physical region increasing with $t$.
This is beneficial since the solution is valid in the vicinity of the 
black hole singularity, so the unphysical region of the spacetime can 
simply be ignored or potentially replaced using a spacetime matching.

In the white hole case, the energy density is non-negative 
everywhere if 
$\dot{R}R$ is non-negative and $\ddot{R}R+\dot{R}^2$ is non-positive, which 
requires $R(t) \sim t^x$ with $0 < x \leq 1/2$, such that the white holes 
expand with time.
With these conditions, $p \geq 0$ and $p \leq \mu$, so all the energy 
conditions are obeyed throughout the spacetime.
The pressure is zero when $R(t) \sim t^{1/2}$, so this is the 
case of pressureless dust.

The $G^0_1$ and $G^1_0$ terms are heat conduction terms, which can arise 
from a null fluid or from energy conduction, either due to the transfer of 
heat through the matter or a non-comoving velocity field that leads 
to a flux of matter relative to the co-ordinates.
The null vector field is $l^a 
=(-1,\pm1,0,0)$, so a null fluid energy-momentum component 
$\tau l^a l_b$ would lead to terms of equivalent magnitude (different 
signs) in $G^0_0$, $G^0_1$, $G^1_0$, and $G^1_1$; however, $G^0_0=0$.
Since $u^au_a=-1$ and
\begin{equation}
u^0u_0 = \frac{T^0_0 - pg^0_0}{\mu + p} = \frac{-p}{\mu+p} \neq -1
,
\end{equation}
while
\begin{equation}
u^0u_0+u^1u_1 = \frac{T^0_0-pg^0_0}{\mu+p} + \frac{T^1_1-pg^1_1}{\mu+p}
= \frac{-\mu-p}{\mu+p} =-1
,
\end{equation}
then there must be a radial velocity field component.
Thus, the heat conduction terms should be interpreted as the flux 
of matter relative to the co-ordinates with $u^1u_1=-\mu/(\mu+p)$.

\section{Finding the horizons}

The new solutions of Section II are non-stationary.
In the case of Schwarzschild, $g_{00}=g^{11}=0$ at $r=2m$;
however, the $g_{00}=g^{11}=0$ surface for the new solutions is at 
$r=2mR^2$, so as $R$ shrinks or grows with time, that surface moves 
to smaller or larger $r$.
With the Schwarzschild metric, the surface $r=2m$ is 
a null surface where null geodesics obey $\mathrm{d}r/\mathrm{d}t=0$, so 
photons at $r=2m$ are held fixed there such that outgoing photons cannot 
move outside the surface in the case of a black hole and ingoing photons 
cannot move inside the surface in the case of a white hole.
With the new dynamic black and white holes it is also true that
null geodesics obey $\mathrm{d}r/\mathrm{d}t=0$ at the surface 
$g_{00}=g^{11}=0$.
However, the surface cannot be the event horizon of the dynamic black 
and white holes, 
since photons at that surface are held at fixed $r$ while the surface 
$r=2mR^2$ shrinks or expands, allowing outgoing photons to escape in 
the black hole case or ingoing photons to become enveloped in the white 
hole case.

If the Kerr-Schild transformation is performed using the geodesic null 
vector field $n^a = (1, \mp 1, 0, 0)$ instead of $l^a = (-1,\pm 1, 0, 0)$,
the spacetime is identical, and $n^a$ can be used to represent the ingoing 
null geodesics of the black holes or outgoing null geodesics of the white 
holes.
Since this geodesic null vector field is preserved under Kerr-Schild 
transformation, the expansion of these geodesics is the same as in Minkowski
space and the divergence is given by
\begin{equation}
n^a_{||a} = n^a_{|a} + \Gamma^a_{ba} n^b = \mp \frac{2}{r}
,
\end{equation}
such that the ingoing null geodesics of the black holes are always converging
and outgoing null geodesics of the white holes are always diverging as expected.
The null vector field representing the opposite-directed null geodesics is 
$k^a = (1/2, \pm 1/2, 0, 0)$ (normalized such that $n^a k_a = -1$), and it is 
not preserved under the Kerr-Schild transformation since
\begin{equation}
\bar{g}^{ab} k_b = g^{ab} k_b - 2 H n^a n^b k_b = k^a + 2 H n^a
.
\end{equation}
The outgoing null geodesics of the black holes and ingoing null geodesics of
the white holes can be represented as
\begin{equation}
\bar{k}^a = (\frac{1}{2}+\frac{2mR^2}{r},\pm \frac{1}{2} \mp \frac{2mR^2}{r},0,0)
,
\end{equation}
such that the outgoing null geodesics of the black holes and ingoing null 
geodesics of the white holes have divergence
\begin{equation}
\bar{k}^a_{||a} = \frac{4mrR\dot{R} \pm r \mp 2mR^2}{r^2}
.
\end{equation}
Thus, unlike the static $\dot{R}=0$ case, the marginal surfaces are not simply at 
$r=2mR^2$.
Since $\dot{R}$ is negative for the black holes and positive for the white holes,
then at $r=2mR^2$ the expansion is negative for the black holes and positive for the 
white holes such that the apparent horizon exists outside $r=2mR^2$ at
\begin{equation}
r= \frac{2mR^2}{1 - 4mR |\dot{R}|}
.
\end{equation}
The expansion of the outgoing null geodesics of the black holes switches
from negative inside this surface to positive outside, and the expansion 
of the ingoing null geodesics of the white holes switches from negative 
outside this surface to positive inside.
Since $R \dot{R}$ varies inversely with $t$ compared with $R^2$, then regions 
that are trapped in the black hole case (or anti-trapped in the white hole case) 
cease to be and the apparent horizon moves inward, asymptotically approaching 
$r=2mR^2$ with time.

It may seem counterintuitive that the apparent horizon is not simply
the $g_{00} = g^{11} = 0$ surface where the outgoing null geodesics of the
black holes and ingoing geodesics of the white holes are instantaneously held 
at fixed areal radius $r$.
However, using the volume element, the co-ordinate volume expansion for a 
spherical shell of infinitesimal thickness is
\begin{equation}
\frac{\dot{V}}{V} = \frac{2mR\dot{R}}{r+2mR^2}
,
\end{equation}
so even when the null geodesics are held at fixed areal radius for an instant, 
they are generally converging in the black hole case and diverging in the 
white hole case due to the volume expansion of the space.

Unlike typical cases of black hole collapses where the trapped region grows
and is contained within an event horizon, since the trapped region is decreasing
an event horizon can only exist inside the apparent horizon.
Since photons can escape the $r=2mR^2$ surface as it moves inward in the black 
hole case, or photons can enter the surface is it expands and envelops them in the
white hole case, this surface cannot generally be the event horizon, and 
the event horizon can only asymptote towards it from within.
The black hole area law also requires that the event horizon
not move to smaller areal radius $r$: since the $2mR^2$ surface shrinks
to smaller $r$ for the dynamic black holes, the only way the event horizon can 
asymptote to it at infinite time is by expanding outward to reach it as a 
fraction of $2mR^2$.
Thus, the aim is to find a surface specified by $r=2mR^2/h$ 
(where $h$ can vary with time) that shrinks or expands at the same rate as 
the null geodesics at that surface,
and such that within it the null geodesics can only move inward or outward 
relative to the shrinking or expanding surfaces in the black hole and white 
hole cases respectively.
It is possible that satisfying this local requirement may somehow differ from
studying the causal structure of the complete spacetime, but 
it is the most reasonable local description that should be equivalent to the 
existence of an event horizon by the usual definition.
The goal is to verify that the spacetimes behave like black and white holes
are qualitatively expected to, rather than strictly showing that they satisfy 
the usual definition of an event horizon, since the time dependence of the 
Kerr-Schild scalar field makes it difficult to produce conformal diagrams of
the spacetimes.

Ignoring the angular components of the line element and looking at where 
it is null yields
\begin{eqnarray}
\mathrm{d}s^2 =  - \mathrm{d}t^2 + \mathrm{d}r^2 + \frac{2mR^2}{r} 
(\mathrm{d}t \pm \mathrm{d}r)^2 \\
0 = - \mathrm{d}t \pm \mathrm{d}r + \frac{2mR^2}{r}(\mathrm{d}t \pm 
\mathrm{d}r) \\
\frac{\mathrm{d}r}{\mathrm{d}t} = \pm \frac{1-2mR^2/r}{1+2mR^2/r} 
.
\end{eqnarray}
The rate at which a surface specified by $h=2mR^2/r$ moves radially is 
given by 
\begin{equation}
\frac{\mathrm{d}r}{\mathrm{d}t} = \frac{4mR\dot{R}}{h}
,
\end{equation}
so equating the rates for the motion of the surface and the motion of 
null geodesics yields 
\begin{equation}
\frac{4mR\dot{R}}{h} = \pm \frac{1-h}{1+h}
.
\end{equation}

\begin{figure}[h]   
\centering
\includegraphics[width=5in]{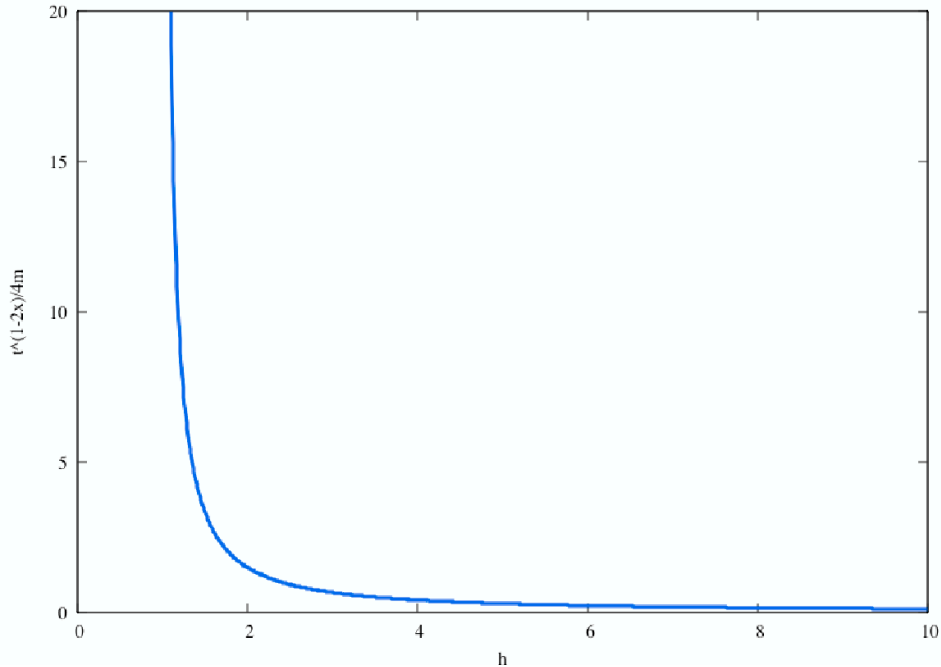}
\caption{Plot of $t^{1-2x}/4m$ versus $h$ for the surface that moves at 
the same rate as the null geodesics in the case of the new dynamic black 
and white holes (with $h=2mR^2/r$). } 
\label{fig1}
\end{figure}

Since $\dot{R}$ is negative for the black hole case and positive for the 
white hole case, then in both cases
\begin{equation}
4mt^{2x-1} \sim 4mR |\dot{R}| = \frac{h^2-h}{h+1}
.
\end{equation}
Since $2x-1$ is non-positive for the black and white hole 
solutions, time increases as a function of $(h+1)/(h^2-h)$.
In figure~\ref{fig1}, a plot of $(h+1)/(h^2-h)$  appears.
It is apparent that $h=\infty$ at $t=0$, which corresponds to $r=0$.
As $t$ approaches infinity, $h$ approaches 1, which corresponds to 
$r=2mR^2$.
Thus, the surface that shrinks or 
expands at the same rate as the null geodesics move
is actually moving outward as a fraction of $2mR^2$ with time, while the 
surface $2mR^2$ shrinks or expands with time.

To determine whether this surface behaves like an event horizon, 
rather than merely temporarily following the null geodesics,
it is necessary to study what happens to null geodesics within it.
Looking at how the $\mathrm{d}r/\mathrm{d}t$ rates are affected going to 
smaller $r$ (larger $h$) at a given moment in time,
\begin{equation}
\frac{\mathrm{d}}{\mathrm{d}h} \frac{\mathrm{d}r}{\mathrm{d}t} = \mp 
\frac{2}{(1+h)^2} 
\end{equation}
for the null geodesics and 
\begin{equation}
\frac{\mathrm{d}}{\mathrm{d}h} \frac{\mathrm{d}r}{\mathrm{d}t} = 
- \frac{4mR\dot{R}}{h^2} 
\end{equation}
for the surfaces.
The null geodesics move inward or outward faster at smaller $r$ for 
the black holes and white holes respectively, while
since $\dot{R}$ is negative for the black holes and positive for the 
white holes, the surfaces move inward or outward slower at smaller $r$.
Thus, within the surface that is shrinking or expanding at the same rate 
as the null geodesics move, photons must move inward or outward faster 
than the interior surfaces move, forcing photons to move toward the black 
hole singularity or toward the white hole surface.
Since the null geodesics within the surface move inward and 
are trapped for all time in the black hole case, the surface appears
to act as a future event horizon.
Since the null geodesics within the surface move outward toward the 
surface in the white hole case, no 
ingoing photons are able to cross the surface and then move inward, 
so the surface appears to act as a past event horizon.

In the case of Vaidya's cosmological black holes \cite{Vai77}, 
repeating the above analysis to try to locate the event horizon 
yields an identical result. 
The reason can be seen most directly by the fact that Vaidya's 
cosmological black hole spacetime is conformally related to the new 
dynamic black holes.
Performing a conformal transformation of Vaidya's cosmological black 
holes to cancel out the scale factor in the seed part of the 
metric yields
\begin{equation}
\mathrm{d}s^2 =  - \mathrm{d}t^2 + \mathrm{d}r^2 + r^2 
(\mathrm{d}\theta^2+\mathrm{sin}^2\theta \mathrm{d}\phi^2) +
\frac{2m}{r[R(t)]^2}(\mathrm{d}t + \mathrm{d}r)^2 
,
\end{equation}
which looks exactly like the metric for the dynamic black holes, except 
that the scale factor is in the denominator of the Kerr-Schild term 
instead of the numerator.
However, the scale factor for Vaidya's cosmological black holes 
grows with time so that $R(t) \sim t^x$ with $x > 0$,
whereas the dynamic black holes have $R(t) \sim t^x$ with $x < 0$, so 
in fact the metrics are the same. 
Thus, the metrics are conformally related, and since conformal 
transformations preserve causal structure, Vaidya's cosmological 
black holes must share any event horizon that the dynamic black 
holes possess.
It should be noted that expansion is not generally conserved under a 
conformal transformation, so the expansion of the outgoing null geodesics 
will differ from that of the new dynamic black holes and be given by
\begin{equation}
\bar{k}^a_{||a} = \frac{r^2R\dot{R}-2m+rR^2}{r^2}
.
\end{equation}
Thus, for expanding universes, the expansion will be positive at the 
$r=2m/R^2$ surface, so the marginal surfaces representing the apparent 
horizon will actually be inside the $g_{tt} = g^{rr} = 0$ surface, 
unlike for the new solutions presented in this paper.

\section{Two-Fluid Solutions}

Performing a Kerr-Schild transformation of Minkowski space with a
scale factor $R(t)$ modifying the usual scalar field used to obtain the 
Schwarzschild metric and with $m=m(u)$ (where $u=t \pm r$) as with Vaidya's 
radiating stellar exterior, the new line element 
\begin{equation}
\mathrm{d}s^2 =  - \mathrm{d}t^2 + \mathrm{d}r^2 + r^2 
(\mathrm{d}\theta^2+\mathrm{sin}^2\theta \mathrm{d}\phi^2) +
\frac{2m(u)[R(t)]^2}{r}(\mathrm{d}t \pm \mathrm{d}r)^2 
\end{equation}
is obtained (with the plus or minus signs corresponding to the black 
or white hole cases respectively).
Looking at the Einstein tensor, the only
non-zero components are
\begin{eqnarray}
G^0_0 = \pm \frac{2R^2m'}{r^2} \\
G^0_1 = - G^1_0 = \frac{4m\dot{R}R}{r^2} + \frac{2R^2m'}{r^2}\\
G^1_1 = \mp \frac{8m\dot{R}R}{r^2} \mp \frac{2R^2m'}{r^2} \\
G^2_2 = G^3_3 = \frac{2m(\ddot{R}R+\dot{R}^2)}{r}
,
\end{eqnarray}
where $m'=\mathrm{d}m/\mathrm{d}t= \pm \mathrm{d}m/\mathrm{d}r$.
Comparing with the case of Section II where $m'=0$,
it is apparent there are now $2R^2m'/r^2$ terms in $G^0_0$, $G^0_1$, 
$G^1_0$, and $G^1_1$ that can be interpreted as a null fluid component 
since the $G^0_0$ and $G^1_1$ terms sum to zero, and the $G^0_1$ and $G^1_0$ 
terms represent heat conduction from the transport of energy as the null 
fluid radially transfers energy.

Thus, this interpretation results in a two-fluid solution consisting of the 
perfect fluid found in Section II combined with a null fluid 
analogous to that of Vaidya's radiating stellar exterior.
In the case of the black hole the null fluid must be ingoing such that the 
singularity accretes mass, and in the case of the white hole the null 
fluid must be outgoing such that the singularity radiates away mass.
Since the additional fluid component is a null fluid, it must satisfy the 
energy conditions, so in combination with the perfect 
fluid component, the solutions should be physical under the same 
conditions as they are in Section II.

\section{Discussion and summary}

Solutions for new non-isolated dynamic black and white holes have been 
found.
The solutions consist of a perfect fluid that is contracting in the black 
hole case and expanding in the white hole case, and if the mass of the 
singularity is allowed to vary across null surfaces there is an 
additional null fluid component that is accreted by the black hole or 
radiated by the white hole.
These exact solutions could serve as simple models of black holes, white 
holes, and stellar exteriors with surrounding matter distributions of 
non-cosmological nature.
While the black hole solutions are only physical within the neighbourhood 
of the singularity, it is possible to match spacetimes together using 
junction conditions, so it is possible these solutions could 
be matched onto another spacetime to make the solution physical throughout.

It is interesting that while having $m$ times a function of time in the 
metric looks analogous to Vaidya's radiating star \cite{Vai43} or Dawood 
and Ghosh's dynamic black holes \cite{Daw04} since they both have the mass 
varying as a function of time, the new dynamic solutions do not merely
yield a null fluid like the radiating star.
This can be explained by the fact that the previous spacetimes have $m$ as a 
function of $u=t \pm r$ , so that $2m/r$ is 
varying between null surfaces, whereas the new dynamic black holes 
have $2m/r$ being scaled by the scale factor between different spacelike 
surfaces. The previous spacetimes are consistent with mass being radiated
away from a white hole or onto a black hole, whereas the 
new dynamic black holes require a different form of mass-energy.

Incorporating a scale factor in the Kerr-Schild 
term of the Schwarzschild metric has a similar effect to performing a 
conformal transformation of Minkowski space in that it introduces 
mass-energy.
A conformal transformation of Minkowski space introduces mass-energy 
since the gravitational influence causes a decelerating expansion of 
space (or equivalently an accelerating shrinking of space as the matter 
falls together backwards in time).
Thus, it makes sense that introducing a scale factor in the Kerr-Schild term 
of the Schwarzschild metric also leads to the introduction of mass-energy,
and it makes sense that this would be inhomogeneous due to the $1/r$ 
dependence of the scalar field.

Presumably the reason the white holes are limited from expanding faster than 
$R^2 \sim t$ is that faster expansions would lead to an acceleration, rather 
than a deceleration, in the expansion, which would not be consistent with 
gravitational attraction of mass-energy.
Since the black holes are shrinking, presumably there should be no problem 
with the mass-energy distribution accelerating together, which would explain 
why they can shrink at any rate, with increasing energy density corresponding 
to a faster decrease in $R$.

While the new solutions are spherically symmetric, the presence of
pressure means they are not examples of Lema\^{i}tre-Tolman-Bondi 
solutions \cite{Lem33,Tol34,Bon47}.
The pressure gradients will exert a force to accelerate the matter.
The pressure goes as $\mp 1/r$, so the pressure gradient is exerting an 
inward force in the black hole case and an outward force in the white 
hole case, meaning the matter will be accelerated toward the black hole and 
away from the white hole.
Since the matter is being accelerated away from the white hole, 
it suggests pressure gradients may be one way to prevent white holes from 
being unstable to recollapsing to form black holes as Eardley has 
suggested they would \cite{Ear74}.

It is interesting that neither the apparent horizon nor the event horizon
seems to coincide with the $r=2mR^2$ surface of the dynamic black and
white holes, yet both asymptote towards it at infinite time.
Due to the decreasing magnitude of the volume expansion of the space, regions 
that are trapped or anti-trapped cease to be, so the apparent horizon 
asymptotically moves inward toward the $r=2mR^2$ surface.
Contrary to usual situations where black holes accrete mass and the apparent
horizon grows to reach an event horizon from within, the apparent horizon 
cannot be contained within an event horizon.
Based on the behaviour of the null geodesics, it appears that there is an
event horizon that prevents null geodesics from escaping the black hole or
entering the white hole, and this horizon asymptotes to the $r=2mR^2$ surface
from within it.

While it was claimed the $g_{tt }= g^{rr} = 0$ surface of Vaidya's 
cosmological black holes \cite{Vai77} is an event horizon, this cannot 
generally be so. 
It has been demonstrated that Vaidya's cosmological black holes have the 
same causal structure as the new dynamic black holes found in this paper, 
and likewise there is a surface within the $g_{tt} = g^{rr} = 0$ surface 
that traps photons for all time, verifying that these cosmological black 
holes do in fact behave as black holes.

\begin{acknowledgments}
Thanks go to Charles Dyer, Charles Hellaby, and 
Amanda Peet (the string theorist, not the actress) for useful 
discussions and comments.
This research was funded by the NRF of South Africa.
\end{acknowledgments}


\newpage

\begin{thebibliography}{99}

\bibitem{Ker65} R. P. Kerr and A. Schild, in \textit{Atti del Convegno 
Sulla Relativita Generale: Problemi Dell'Energia E Onde Gravitazionale}, 
edited by G. Barbera (Comitato Nazionale per le 
Manifestazione celebrative, Florence, 1965) p. 222.
\bibitem{Ste03} H. Stephani, D. Kramer, M. A. H. MacCallum, 
C. Hoenselaers, and E. Herlt, \textit{Exact Solutions to Einstein's Field 
Equations}, 2nd Ed. (Cambridge U. Press, Cambridge, 2003) p. 485.
\bibitem{Daw04} A. K. Dawood and S. G. Ghosh, Phys.\ Rev.\ D \textbf{70}, 
104010 (2004).
\bibitem{Vai43} P. C. Vaidya, Current Science \textbf{12}, 183 (1943); 
Gen.\ Relativ.\ Gravit. \textbf{31}, 119 (1999). 
\bibitem{Sal03} M. Salgado, Class.\ Quantum Grav. \textbf{20}, 4551 (2003).
\bibitem{McV33} G. C. McVittie, Mon.\ Not.\ R.\ Astron.\ Soc. \textbf{93}, 
325 (1933).
\bibitem{Vai77} P. C. Vaidya, Pramana \textbf{8}, 512 (1977).
\bibitem{Tha81} S. N. G. Thakurta, Indian J.\ Phys. \textbf{55B}, 304 
(1981).
\bibitem{Sul05} J. Sultana and C. C. Dyer, Gen.\ Relativ.\ Gravit. 
\textbf{37}, 1349 (2005).
\bibitem{McC06} M. L. McClure and C. C. Dyer, Class.\ Quantum Grav. 
\textbf{23}, 1971 (2006).
\bibitem{Ein45} A. Einstein and E. G. Straus, Rev.\ Mod.\ Phys. 
\textbf{17}, 120 (1945).
\bibitem{Har94} J. H. Harper and C. C. Dyer, \textit{Tensor Algebra 
with REDTEN} (U. of Toronto, Scarborough, 1994).
\bibitem{Ash04} A. Ashtekar and B. Krishnan, Living Rev.\ Relativity 
\textbf{7}, 10 (2004).
\bibitem{Hay94}	S. A. Hayward, Phys.\ Rev.\ D \textbf{49}, 6467 (1994).
\bibitem{Wal91} R. M. Wald and V. Iyer, Phys.\ Rev.\ D \textbf{44}, 3719 
(1991).
\bibitem{Sch06} E. Schnetter and B. Krishnan, Phys.\ Rev.\ D \textbf{73}, 
021502 (2006).
\bibitem{Wal84} R. M. Wald, \textit{General Relativity} (U. of Chicago 
Press, Chicago, 1984) p. 219.
\bibitem{Bon47} H. Bondi, Mon.\ Not.\ R.\ Astron.\ Soc. \textbf{107}, 410 
(1947); Gen.\ Relativ.\ Gravit. \textbf{11}, 1783 (1999).
\bibitem{Lem33} G. Lema\^{i}tre, Ann.\ Soc.\ Sci.\ Bruxelles 
\textbf{A53}, 51 (1933); Gen.\ Relativ.\ Gravit. \textbf{29}, 641 (1997).
\bibitem{Tol34} R. C. Tolman, Proc.\ Nat.\ Acad.\ Sci.\ U.\ S.\ A. 
\textbf{20}, 169 (1934); Gen.\ Relativ.\ Gravit. \textbf{29}, 935 (1997).
\bibitem{Ear74} D. M. Eardley, 1974 Phys.\ Rev.\ Lett. \textbf{33}, 442 
(1974).

\end{thebibliography}
\end{document}